\documentclass[fleqn,reprint,groupedaddress,amsmath,amssymb,aps,prx,floatfix,longbibliography]{revtex4-1}
\usepackage{graphicx}
\usepackage{amsmath}
\usepackage{multirow} 
\usepackage{physics} 
\usepackage{xcolor}
\graphicspath{{Figures/}}

\begin{document}

\title{Protocol for autonomous rearrangement of cold atoms into low-entropy configurations}

\author{Matthew A. Norcia}
\email[E-mail: ]{matthew.norcia@uibk.ac.at }
\affiliation{
    Institut f\"{u}r Quantenoptik und Quanteninformation, \"Osterreichische Akademie der Wissenschaften, Innsbruck, Austria
}
\begin{abstract} 
The preparation of low-entropy starting conditions is a key requirement for many experiments involving neutral atoms.  
Here, we propose a method to autonomously assemble arbitrary spatial configurations of atoms within arrays of optical tweezers or lattice sites, enabled by a combination of tunneling and ground-state laser cooling.  In contrast to previous methods, our protocol does not rely on either imaging or evaporative cooling.  This circumvents limitations associated with imaging fidelity and loss, especially in systems with small spatial scales, while providing a substantial improvement in speed relative to evaporative approaches.  These features may make it well-suited for preparing arbitrary initial conditions for Bose-Hubbard or Rydberg interacting systems.

\end{abstract}

\date{\today}

\maketitle

\section{Introduction}

Microscopic control of neutral atoms has lead to a recent explosion of interest for applications in quantum simulation, quantum information processing, and quantum enhanced metrology \cite{gross2017quantum, browaeys2020many, weiss2017quantum, pohl2014}.  
Achieving configurations of large numbers of atoms with well-defined positions and motional states is a key capability at the forefront of modern experiments.
Currently, two main approaches are used to accomplish this: In the first, evaporative cooling is used to remove entropy from the atomic system, after which point the atoms may be adiabatically loaded into a desired potential landscape such as an optical lattice \cite{gross2017quantum}.  
Entropy redistribution techniques can then be used to further reduce the entropy of a sub-region \cite{Chiu2018, kantian2018dynamical, yang2020cooling}.  
In the second approach, atoms are stochastically loaded into the tightly confining potential of optical tweezers or lattice sites, with light-assisted collisions leading to either zero or one atom in each site \cite{schlosser2001sub}.  The atoms are then rearranged into the desired configuration based on information gained from imaging \cite{Weiss2004, miroshnychenko2006atom, kim2016situ, Barredo2016, Endres2016}.  Of particular relevance for this proposal are recent experiments where atoms are also laser-cooled to their motional ground states \cite{Kumar2018}.


While highly successful, these two existing approaches each have limitations.  Evaporation is typically slow, leading to long experimental cycle times.  This can be problematic for quantum simulation experiments, where large numbers of experimental trials are required, and in the context of metrology, where long cycle times degrade the stability of atomic clocks or other sensors \cite{Dick1987}.  
Approaches based on measurement and rearrangement can provide a substantial advantage in terms of speed, as laser cooling replaces evaporation as the means of entropy removal.  However, the resulting level of entropy can be limited by imaging fidelity and loss.  Further, these approaches require an imaging system capable of localizing and manipulating atoms at the single-site level, which has so far prevented the application of rearrangement protocols to systems with sub-micron-scale lattices, desirable for tunneling.  

In this work, we propose an alternate approach that does not require either evaporation or imaging.  
Like the imaging-based approach, our method starts with an array of stochastically loaded atoms to be rearranged into a desired configuration.  However, rather than relying on imaging to determine the initial site occupations, atoms are rearranged in an autonomous manner. For clarity, we will leave further motivation and connections to related concepts such as Thouless pumping, algorithmic cooling and autonomous stabilization for later (Section~\ref{sec:discussion}) and proceed immediately with a description of our proposed mechanism.

\section{Protocol Overview}

\begin{figure}[!htb]
\includegraphics[width = \linewidth, ]{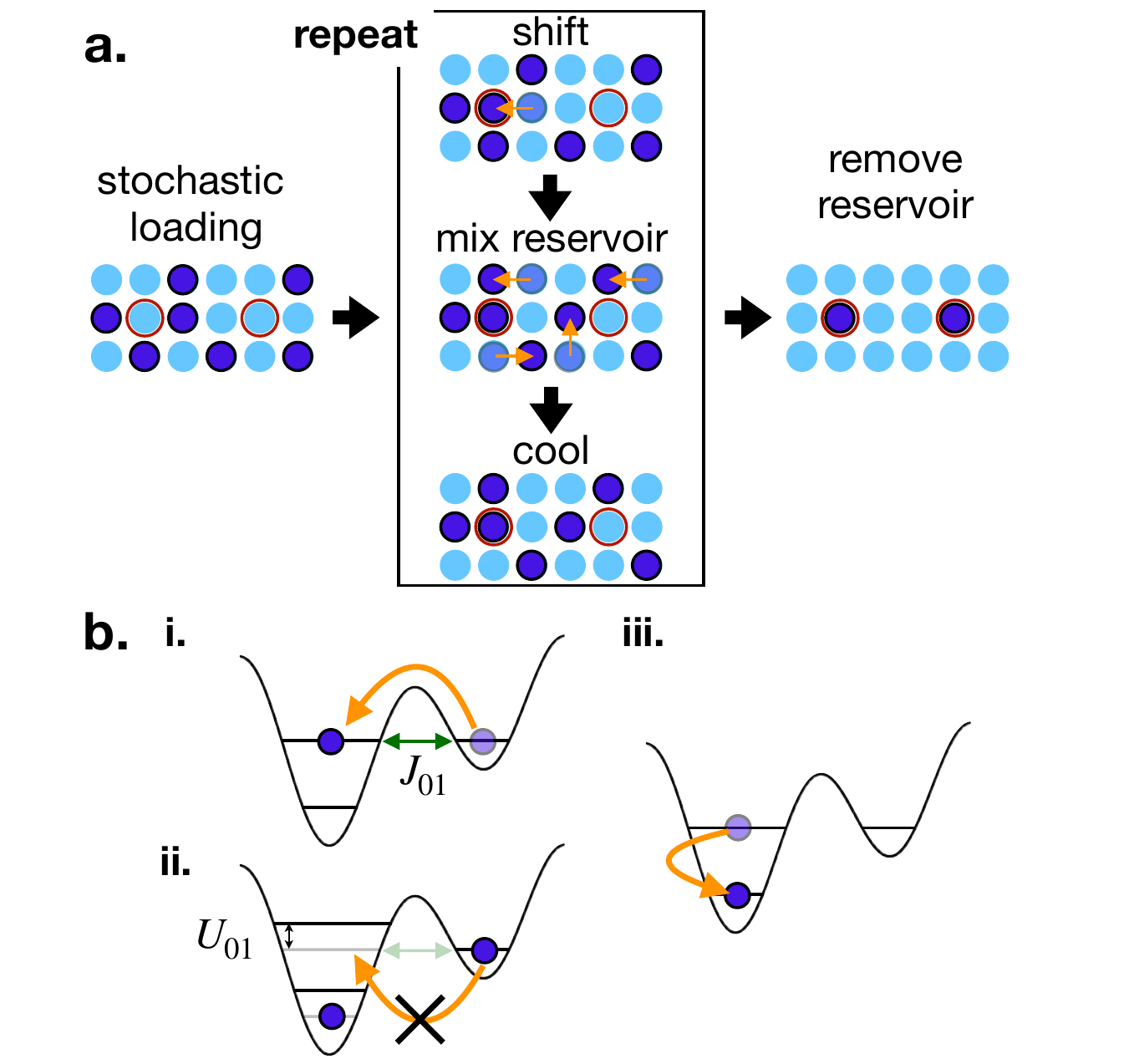}
\caption{  
\textbf{a.} Conceptual overview of method. An array of lattice sites, consisting of target (red circles) and reservoir (no circles) sites is initially loaded in a stochastic manner.  After subsequent cycles of directional tunneling into target sites and between reservoir sites, interleaved with dissipative cooling steps, target states within the array are occupied. Atoms remaining in the reservoir sites are then removed, leaving a tailored low-entropy configuration.
\textbf{b.} Basic building block of implementation.  \textbf{i.} An optical potential consisting of two neighboring wells is imbalanced to create a tunneling resonance between the ground state of the reservoir site (right well) and the first motionally excited state of the target site (left well).  An atom occupying the reservoir site will then tunnel into an empty target site.  \textbf{ii.} If an atom already occupies the target site, contact interactions shift the tunneling resonance, preventing double occupation of the target site.  \textbf{iii} After tunneling, ground-state laser cooling is applied to return atoms to the ground state of their respective sites, providing dissipation and preventing future tunneling away from the target site.  
}
\label{fig:overview}
\end{figure}

The goal of our protocol is to transfer atoms into specific ``target" sites of an optical potential from stochastically loaded ``reservoir" sites by repeated application of irreversible ``shift" operations, after which point atoms remaining in reservoir sites are removed (Fig.~\ref{fig:overview}a).  
The shift operation can be understood in a simple subsystem consisting of a single target site and a single reservoir site.  If initially the target site is empty and the reservoir site is occupied by an atom, the atom is transferred from the reservoir site to the target site.  For all other configurations, the state remains unchanged.  


Our proposed implementation for the shift operation is shown in Fig.~\ref{fig:overview}b.  We assume that either zero or one atom initially occupies the motional ground state (denoted by $n=0$) of each of two optical potential wells, defining a reservoir and target site.  Excited motional states are assumed to be empty.  The $n=0$ state of the reservoir well is then brought onto resonance with the first excited motional state in the target well ($n=1$), allowing the atom to tunnel between the two wells at a rate that we label $J_{01}$.  The potential could either be held in this resonant configuration for a time $\pi/J_{01}$, or the relative depth of the wells could be adiabatically ramped through resonance to gain insensitivity to offset errors at the expense of reduced speed.  In either case, an atom occupying the reservoir site will tunnel into the target site.  

In order to prevent double occupation of the target site, the contact interactions $U_{01}$ associated with atoms in the $n=0$ and $n=1$ states of the same site must exceed the tunneling rate  $J_{01}$.  This ensures that tunneling from the reservoir to the target site is not resonant if both sites are initially occupied, so the atom in the reservoir well remains in place.  To make the process non-reversible -- to prevent the atom from tunneling out of the target site on a later cycle --  the shift operation concludes by subjecting all atoms to ground-state cooling. 


In general, a target site can be coupled to several reservoir sites.  In this case, one of the atoms (if present) in any of the reservoir sites may tunnel into the target site.  Unlike the two-site case, this process is inherently probabilistic, depending on the relative degeneracy of the configurations with occupied and unoccupied target sites (see appendix~\ref{sec:app_sims} for details).  However, by sufficient repetition of the shift protocol, the target sites can still be loaded with high probability.  



\begin{figure}[!htb]
\includegraphics[width=\linewidth, ]{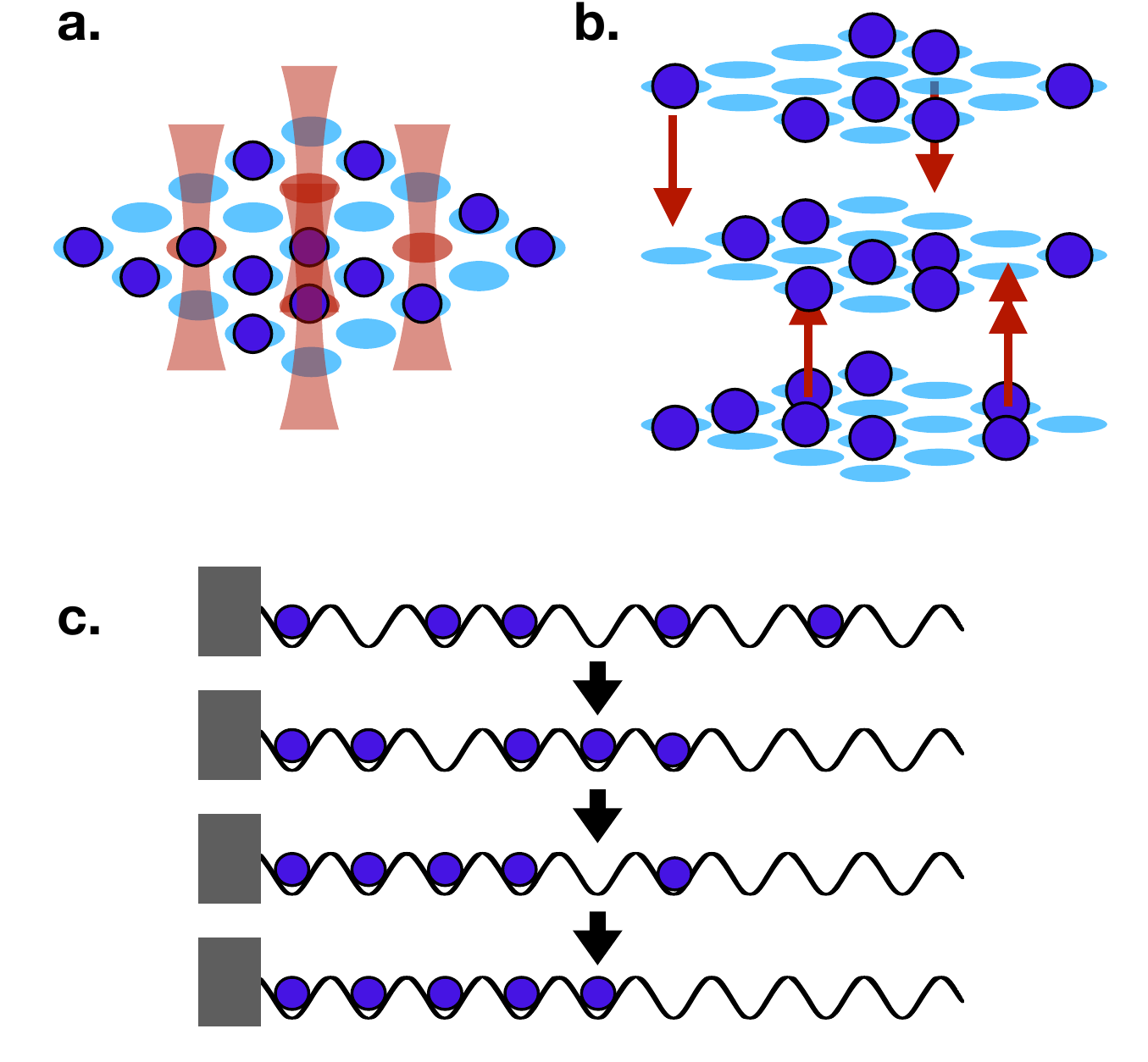}
\caption{Applications to different geometries.  \textbf{a.} Target sites can be defined within a two-dimensional lattice by applying local optical potentials to these sites, for example with optical tweezers.  Shift operations transfer atoms into target sites from neighboring reservoir sites.  
\textbf{b.}  The same concept can be applied to the filling of layers within a three-dimensional lattice.  Here, shift operations are applied between layers, and tunneling within layers enables reservoir atoms to find empty target sites.  
\textbf{c.} A continuous region can be filled by repeatedly shifting atoms towards a potential barrier, represented by the gray region on the left of the array.   
In all protocols, atoms are removed from the reservoir sites after filling of the target sites or layers is complete (see appendix~\ref{sec:app_removal})
}
\label{fig:protocols}
\end{figure}

\begin{figure}[!htb]
\includegraphics[width=\linewidth, ]{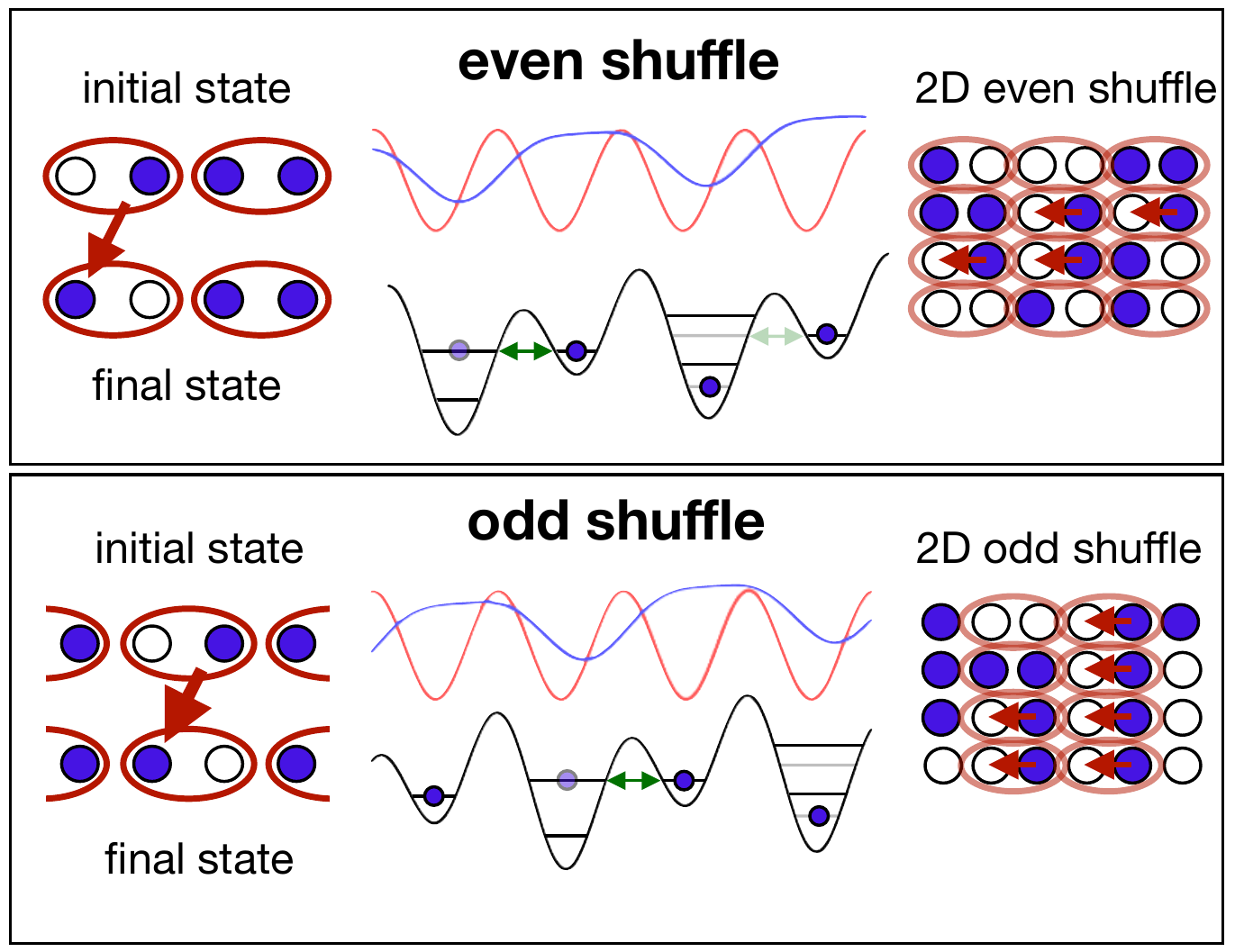}
\caption{Protocol for filling uniform regions without additional dimensions.  By alternately shifting even and odd pairs of atoms, atoms are shuffled to one side, where they bunch up against a barrier (barrier not pictured).  The shift operations are implemented in parallel by superimposing an optical lattice (red lines) with an auxiliary potential (blue lines) that both creates an imbalance between neighboring wells and pushes down the barrier in the combined potential (black lines), enabling rapid tunneling.  The phase of the additional potential is shifted relative to the lattice to shift even (upper) or odd (lower) pairs.  Right:  The shuffle operation can easily be extended to two dimensions simply by applying shuffle operations to many rows in parallel.  Shuffling of each row is independent of others.  }
\label{fig:shuffle}
\end{figure}

\section{Specific implementations}

Variations on this simple building block can be used in different geometrical configurations, for example to fill arbitrary sites within a two-dimensional plane, or to fill target planes within a three-dimensional lattice.  It can also be used in more complex rearrangement algorithms, for example to fill contiguous regions within one- or two-dimensional lattices.  In this section we briefly describe three such variations before turning to a more detailed analysis of their performance.

The conceptually simplest application of our rearrangement protocol can be used to fill arbitrary target sites of a two-dimensional lattice, as illustrated in Fig.~\ref{fig:protocols}a.  In this approach, a background lattice defines a two-dimensional plane of sites among which the atoms can tunnel, and an auxiliary potential (generated using optical tweezers or superlattices), is applied to the target sites.  All other sites are considered reservoir sites. The shift operation is implemented by increasing the depth of the auxiliary potential, adiabatically ramping the energy of the $n=1$ state of the target sites across the $n=0$ states of the reservoir sites.  Importantly, the reservoir sites are assumed to be degenerate in energy, so atoms can also tunnel between them.  This serves the important function of mixing the reservoir atoms so that if a given target site initially lacked neighboring atoms, it may attain them in future cycles.  During the ground-state cooling phase, the depth of the background lattice is increased, localizing each atom on its lattice site, the auxiliary potential decreased, and all atoms cooled to $n=0$.

The same concept can be used to fill target planes instead of target sites, as shown in Fig.~\ref{fig:protocols}b.  In this case, the background lattice is three-dimensional, and the auxiliary potential shifts the energy of entire planes.  The plane shifted to the lowest energy becomes the target plane, and its neighbors the reservoir.  The protocol then proceeds as before, with atoms repeatedly shifted into the target plane.  Again, we assume that the reservoir sites in a single plane are degenerate, so tunneling within the planes serves to randomize the location of atoms, enabling high-probability loading of the target plane.  


The two preceding variations rely on a high connectivity of target sites to reservoir sites, and of reservoir sites to each other.  This can be limiting if one wishes to prepare a state that is uniformly filled without using additional lattice dimensions.  To get around this limitation, a procedure of repeated directional shifts can be used to shuffle atoms into a target region of a lattice, as shown in Fig.~\ref{fig:protocols}c.  This protocol fundamentally operates in a single dimension, but can be used to generate two- or three-dimensional regions with simultaneous application to all rows in the array.  


With this in mind, we consider a single row within a three-dimensional optical lattice, which is tightly confining in the directions orthogonal to its length.
To create a uniformly filled section of the lattice, a potential barrier is applied to one end of a desired region (here,the left end), and shift operations are applied simultaneously to all non-overlapping pairs of lattice sites with an even index of the left well (even pairs), alternating with shift operations applied to all of the pairs with an odd index of the left well (odd pairs), as shown in Fig.~\ref{fig:shuffle}.  We label a pair of such shift operations a shuffle.  Repeated shuffles cause the atoms to move to the left, until they encounter either the barrier or other atoms, at which point further tunneling is suppressed either by the barrier or interaction potential.  After this point, atoms beyond a certain distance from the barrier can be removed from the sample, leaving a uniformly filled region of the lattice (appendix~\ref{sec:app_removal}).

Because the shuffling operations can be applied to all rows of a 2D array simultaneously and independently, we expect the characteristic time and error rates when filling an array of $M$ rows and $N$ columns to be the same as those for a single one-dimensional array of length $N$.  This feature represents a key strength of this technique -- while filling a single one-dimensional array takes a relatively large number of steps, because the atoms are only moved a single site at a time, the shift operations can be applied in parallel to all atoms at the same time, leading to the favorable scaling with respect to total number of atoms in the array, especially when $M$ exceeds $N$.  



\section{Protocol Performance}

\begin{figure}[!htb]
\includegraphics[width=\linewidth, ]{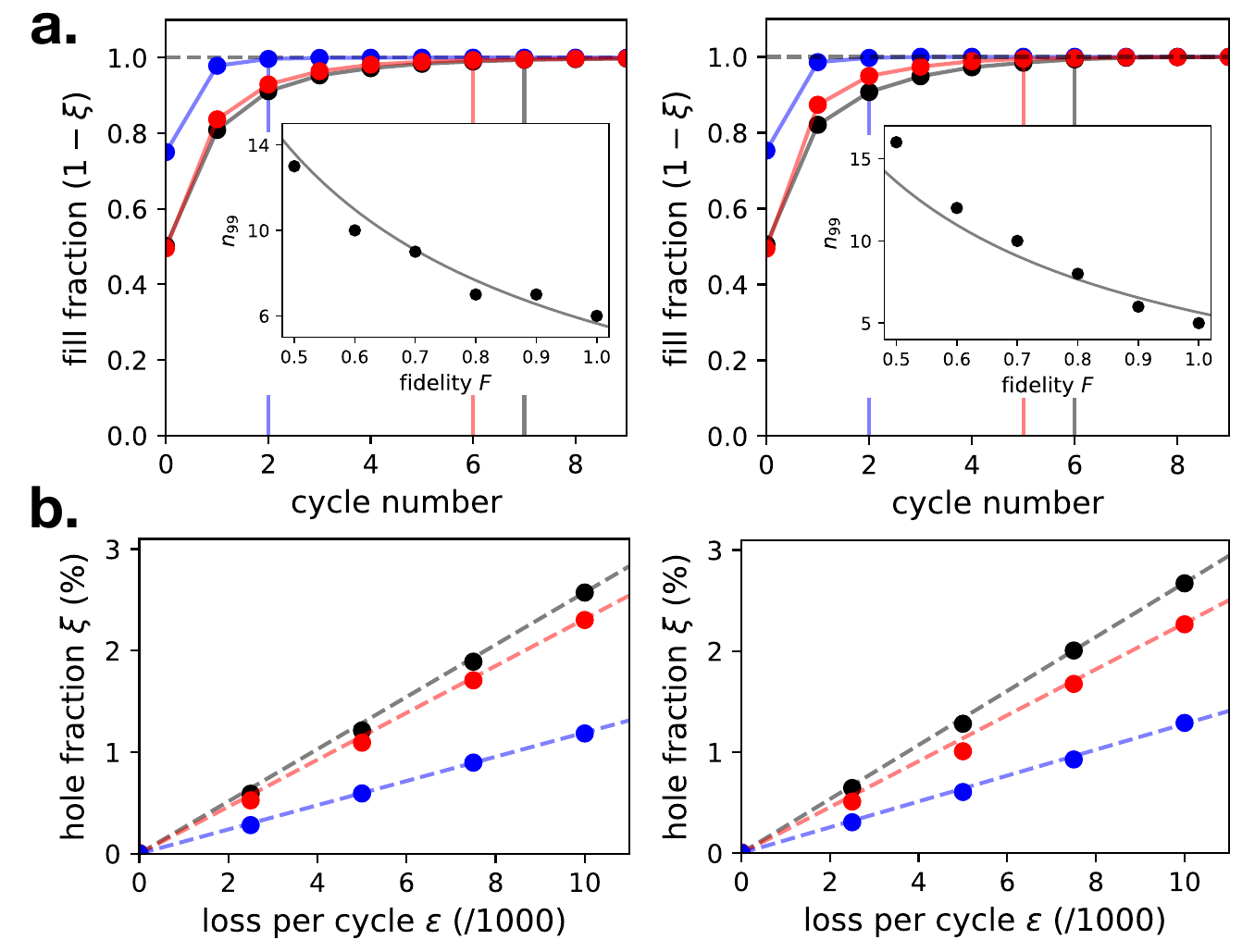}
\caption{Effects of errors on target site (left column) and layer filling (right column) protocols. Red (black) points represent 50\% initial filling, with $F = 1,\ (0.9)$.  Blue points represent 90\% intitial filling, with $F = 1$. \textbf{a.} Filling fraction $1-\xi$ versus number of cycles.  Vertical lines represent the number of steps $n_{99}$ required to reach 99\% filling for each condition.  Loss not included.  Inset: effect of tunneling fidelity on $n_{99}$.  For moderate infidelities, the effect is minor.  Gray line is intuitive approximation discussed in text.  \textbf{b.} Minimum hole fraction $\xi$ fraction versus per-cycle loss, for same conditions as (a).   }
\label{fig:errors}
\end{figure}

The effectiveness of our protocol can be quantified in different situations by the number of cycles required to produce a desired filling (which would dictate experimental cycle times), and by the error rate in the final configuration.  We quantify the former in terms of $n_{99}$, the typical number of cycles to reach 99\% filling, as such a filling would enable assembly of perfect arrays of sizes that can be difficult to simulate classically.  We define the error rate $\xi$ as the minimum achievable fraction of empty target sites.  

Both $n_{99}$ and $\xi$ may be limited by both fundamental aspects of the protocol and by experimental imperfections.  Fundamental limits include the probabilistic distribution of atoms in reservoir sites, and the probabilistic nature of tunneling into a target site that has more than one neighboring reservoir site.  Experimental imperfections can be broken into two categories:  fidelity errors and loss errors.  Fidelity errors represent imperfect tunneling (for example due to imperfections in the optical potential or imperfect adiabatic transfers), leading to a probability of desired tunneling events reduced by a factor $F$ below their idealized value.  Loss errors, quantified by the per-cycle atom loss probability $\epsilon$, represent atom loss from the system due to effects such as collisions with background gas due to imperfect vacuum, double-occupancy of a lattice site, or imperfect ground state cooling.  Detailed estimates for experimentally achievable fidelities and loss rates can be found in appendix~\ref{sec:app_params}, indicating that fidelities $F \geq 0.9$ and loss $\epsilon \leq 0.01$ should be attainable.  

For all variations of our protocol, we find through simulation that for small loss rates and infidelity, the minimum occupancy error $\xi \simeq \alpha \epsilon/F$. $\alpha$ then provides a figure of merit that can be used to assess the performance of a given protocol, target state, and initial filling fraction.  Intuitively, $\alpha/F$ is related to the number of cycles required to repair a defect caused by loss.  Because $\epsilon$ must be kept small for any useful application of our protocol, the effect of loss is minor in determining the timescale required to reach a quasi-equilibrium condition, and we compute $n_{99}$ (which is of course not a good metric for conditions where $\xi \geq 0.01$) in the absence of loss. 

We estimate the performance of our protocols using a classical simulation of the whole system with with parameters informed by master-equation-based simulations performed on much smaller subsystems, details of which are provided in appendix~\ref{sec:app_sims}.  We break each filling cycle into three steps:  a filling step in which atoms may tunnel into an unoccupied target site from neighboring reservoir sites, a mixing step where the atoms are allowed to rearrange between reservoir sites (where applicable), and a loss step.  In the experiment, filling and mixing could occur simultaneously, and be interleaved with ground state cooling (which is not represented in the simulation as we explicitly track only site occupation, but include the effects of imperfect cooling as effective loss).

To benchmark the performance of our first protocol variation, we consider a specific target state -- a grid with quarter filling of the base lattice, as shown in Fig.~\ref{fig:protocols}a.   When the filling fraction of the target state is much lower than the typical filling of stochastically loaded lattices, there is a high probability of having enough atoms to fill the target sites for systems of at least moderate size.  The filling process for such a target state is shown in Fig.~\ref{fig:errors}a.  In the absence of imperfections, we find that $n_{99} = 6$ for 50\% initial filling.  Very roughly, this indicates that on each cycle, the chance that a target site remains unoccupied is reduced by a factor or about 2 (though this factor changes slightly over the filling process as atoms are moved to the target sites, reducing the density of the reservoir).  The effect of imperfect tunneling fidelity is to reduce this factor, leading to slower filling.  For half-filling, we find that an intuitively derived formula: $n_{99} = Log_{1-P}(0.02)$, where $P = 0.5 F$ represents the chance of filling an unfilled target site on any cycle, provides good agreement with the results of our simulation.  

The filling of layers is limited by conceptually similar factors to the filling of target sites.    However, the number of neighbors connected to each target site, and the connectivity of the reservoir sites are not the same for the two geometries, leading to quantitative differences. Further, because the performance of these protocols benefits from a high density of reservoir atoms, the layer filling configuration presents an opportunity: atoms can be brought into the reservoir planes that neighbor the target plane from farther away planes.  This leads to a higher density of atoms in reservoir planes that neighbor the target plane, allowing for more rapid filling of vacancies in the target plane.  This both reduces the number of cycles $n_{99}$ required to fill the target plane, and allows errors associated with loss to be repaired more quickly, reducing the error rate $\xi$.  

As a specific example, we consider a implementation based on five layers, numbered one through five, that are initially stochastically loaded with half filling, with plane number three the target plane.  First, shift operations are applied simultaneously from plane one to two, and from five to four.  This increases the density of atoms in planes two and four.  After this, repeated shift operations are applied into plane three from both of its neighbors, until the desired filling fraction is reached.  

The filling performance of this protocol is shown in figure \ref{fig:errors}b, along with the same prediction for the effects of imperfect fidelity as for the target sites.  In this case, we find $n_{99} = 7$ for perfect fidelity and 50\% initial filling, and that for lower fidelities our intuitive prediction slightly underestimates the number of cycles required. This is because fewer atoms are transferred from planes one and five into the central three in the first cycle, reducing the density of available reservoir atoms.  We note that this quantitative similarity with the target states protocol is specific to the example configurations that we chose, and does not represent a general property of the two approaches.

For both benchmark cases described above (for target sites and target layers with 50\% initial filling), we find from simulation that $\alpha \simeq 2.25$ (Fig.\ref{fig:errors}c, d, red points).  This number roughly reflects the inverse of the probability that a defect is filled on a given filling cycle.  It is influenced both by the availability of atoms in neighboring reservoir sites, and the inherently probabilistic nature of filling when a target site is coupled to multiple reservoir sites (see appendix~\ref{sec:app_sims}).  As mentioned above, these numbers can be dramatically improved by increasing the density of atoms in the reservoir states (Fig.\ref{fig:errors}c, d, blue points), either by shifting atoms into those sites, or through protocols that can lead to filling fractions significantly above 50\% \cite{Grunzweig2010, lester2015, brown2019gray}.

\begin{figure}[!htb]
\includegraphics[width=\linewidth, ]{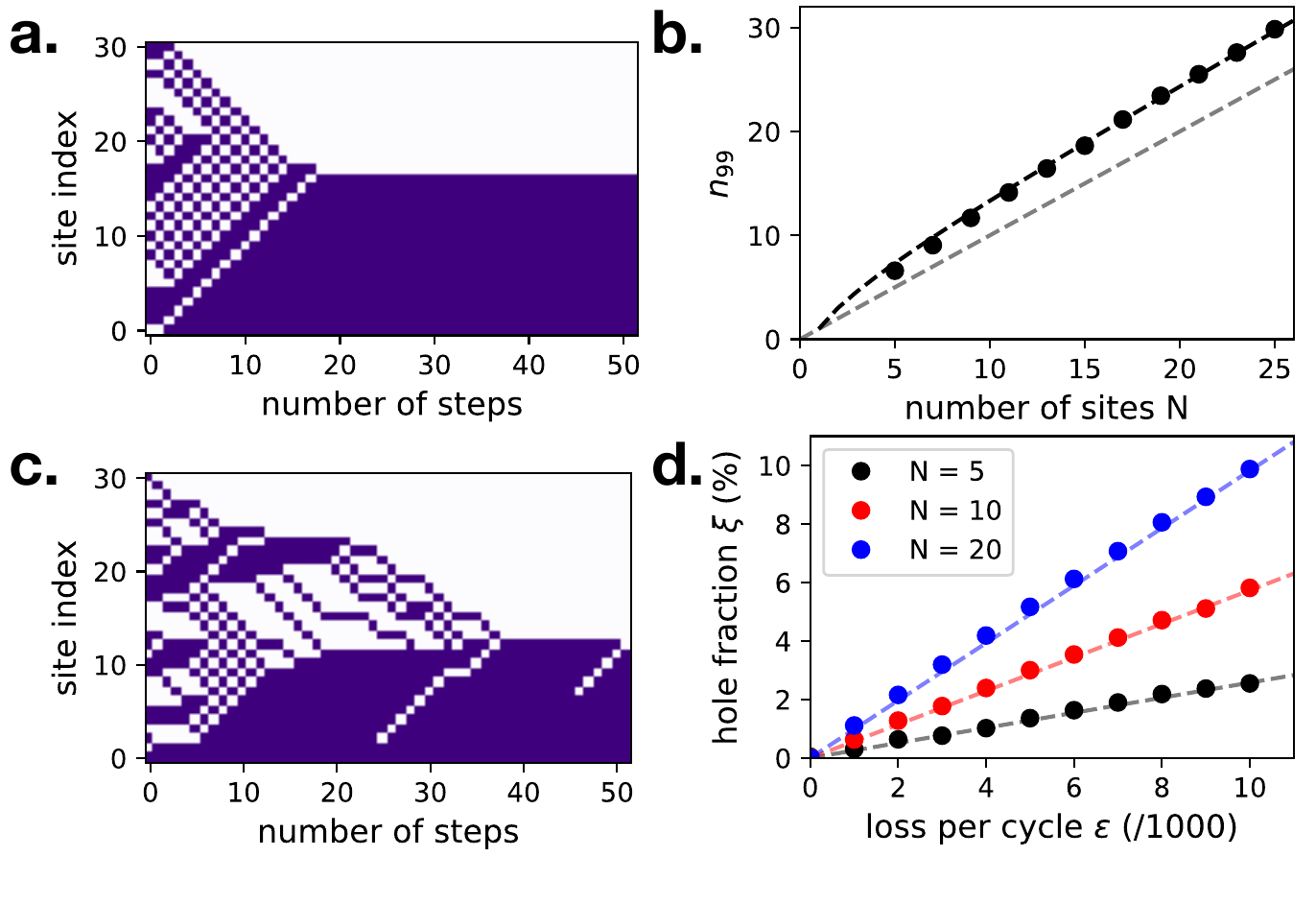}
\caption{Shuffling performance in idealized case and with imperfect fidelity and loss.  \textbf{a.} Example of lattice site occupation versus number of shuffling cycles, with perfect tunneling and no loss.  Each cycle consists of either an even or an odd shift.  After an initial traffic jam, the atoms shuffle at a rate of one site per step until they are blocked by other atoms.  \textbf{b.} Number of cycles required to reach 99\% filling ($n_{99}$) for different target region lengths $N$.  Dashed black line is the prediction from the text, and dashed grey line is $N$.  \textbf{c.} Example of lattice site occupation versus number of shuffling cycles in the presence of finite shift fidelity ($F = 0.9$) and loss per cycle($\epsilon = 0.01$).  
\textbf{d.} Prevalence of holes in the final array versus loss $\epsilon$ and $N$.  }
\label{fig:shuffle_performance}
\end{figure}

The filling process of a one-dimensional sub-array under our shuffling protocol is shown in Fig.\ref{fig:shuffle_performance}.  In the absence of errors (Fig.\ref{fig:shuffle_performance}a, b), and starting from an initial randomly half-filled array long enough to contain at least $N$ atoms with high probability, we make an informed guess that the typical number of shuffles required to fill a sub-array of length $N$ is well approximated by $n_{99} = N + C(N)$, where $C(N)$ is a logarithmic correction factor. 
This guess is motivated by thinking of holes being shuffled out of the sub-array --- each hole can move up to one sites per cycle (either even or odd shuffle), which contributes the linear term.  However, a hole cannot begin to move until there is an adjacent atom to fill it, so if we start with several adjacent holes, the one that must move the farthest does not begin to move until its neighbors have been filled.  Because the longest typical number of adjacent holes within the sub-array should scale with $\log_2(N)$ \cite{schilling1990longest}, we expect this to set the scale of $C(N)$, and find empirically that $C(N) = \log_2(N)$ indeed provides reasonable agreement with simulation (Fig.~\ref{fig:shuffle_performance}b).

In the shuffle protocol, the per-site error rate increases with the length of the region being filled.  For an target region of $N$ by $M$ lattice sites, where shuffling is performed in the direction whose size is $N$, we find from simulation that the achievable error rate is given approximately by $\xi \simeq N \epsilon/2F$, corresponding to $\alpha = N/2$.  Intuitively, this can be understood because a hole in the target region must be moved $N/2$ sites on average, requiring $N/2F$ shuffling cycles (Fig.\ref{fig:shuffle_performance}c, d).  Note again that while the performance scales unfavorably with $N$, the number of columns in the target region, it is independent of the number of rows $M$.  As a benchmark case, given per-shift losses of $\epsilon = 0.01$, one could expect to achieve perfect filling of a five-by-five array in approximately half of attempts.  In experiments where perfect arrays are not needed (for example, for those targeting entanglement-enhanced metrology \cite{van2020impacts}), the system size could be dramatically increased.

\section{Discussion and outlook}\label{sec:discussion}

So far, we have described several ways in which low-entropy arrays of atoms may be autonomously assembled from stochastically populated initial conditions, potentially combining advantages of evaporative techniques, and those based on measurement and rearrangement.  Because entropy is removed through the cooling process, rather than imaging, our techniques are suitable for systems where imaging may have limited fidelity or be accompanied by loss, for example in atomic species with complex level structures or in systems with small lattice spacings.  This may make these techniques especially suitable for systems where preparing atoms in the motional ground state of a short wavelength optical lattice is already required, such as those seeking to explore Hubbard physics.  Further, because the definition of the target sites is can be chosen at will (up to certain constraints), this approach could be useful for preparing initial conditions for Hubbard-regime experiments that require or benefit from arbitrary, non-uniform starting conditions, such as sampling problems \cite{muraleedharan2019quantum} or quantum walk experiments \cite{kempe2003quantum}, or for simulators of spin-models \cite{browaeys2020many, weiss2017quantum}, where interactions can be mediated by Rydberg interactions with range larger than the lattice spacing.

There are some similarities between our method and previously proposed and demonstrated methods for algorithmic cooling, in that both rely on contact interactions between atoms in a lattice to allow or prevent coherent transitions that ultimately enable the reduction of filling errors \cite{rabl2003defect, popp2006ground,bakr2011orbital,tichy2012shaking}.  However, algorithmic cooling relies on overfilling the lattice and then removing the excess atoms, so typically requires an evaporatively prepared sample to begin with.    In contrast, because our method works by irreversibly moving atoms towards the target sites, it is specifically suitable for the relatively sparse initial fillings typically associated with direct laser cooling into micron-scale optical potentials.  
The directional tunneling employed here is similar to the topological charge pumping employed in a Thouless pump, which has been demonstrated to be highly robust to experimental imperfections \cite{romero2007quantum, lohse2016thouless, nakajima2016topological, koepsell2020robust}.  However, the addition of ground-state cooling makes the pumping irreversible, enabling the compression of particles in the sites of an optical lattice and the removal of spatial entropy.

Our scheme also bears resemblance to autonomous quantum stabilizers for bosonic systems, for example those used to stabilize highly correlated photonic states in arrays of superconducting qubits \cite{ma2017autonomous, ma2019dissipatively}.  With suitable modifications, the basic ideas behind our approach could also be used to stabilize analogous atomic systems against the effects of atom loss, and to provide a well-controlled means of environmental coupling.  This will be a direction of future exploration, and may benefit from the use of even narrower transitions, such as the clock transitions in alkaline earth atoms.  In a very recent proposal \cite{Sharma2021}, a related mechanism is explored to autonomously assemble and stabilize atoms within optical lattices, with dissipation provided by interaction with an atomic bath, rather than laser cooling as is used here.  If extended to molecules in tweezers or optical lattices \cite{chotia2012long, anderegg2019optical, cairncross2021assembly}, variations on the method proposed here could be highly advantageous to increase the filling fractions and reduce disorder, as nondestructive imaging of molecules is challenging.  









\section{Acknowledgements}
M.A.N. thanks Adam Kaufman and Hannes Pichler for insightful discussions and feedback on the manuscript.  
This project has received funding from the European Union’s Horizon 2020 research and innovation programme under the Marie Skłodowska‐Curie grant agreement No 801110 and the Austrian Federal Ministry of Education, Science and Research (BMBWF). It reflects only the author's view, the EU Agency is not responsible for any use that may be made of the information it contains
\clearpage

\bibliography{references.bib}

\clearpage
\setcounter{section}{0}
\setcounter{equation}{0}
\setcounter{figure}{0}
\setcounter{table}{0}
\setcounter{page}{1}
\makeatletter
\renewcommand{\theequation}{S\arabic{equation}}
\renewcommand{\thefigure}{S\arabic{figure}}
\appendix
\section{Experimental Implementation}\label{sec:app_implement}
The success of this method relies on strong contact interactions between atoms to prevent multiple-occupancy errors, and a level structure compatible with high-fidelity ground-state laser cooling.  One atomic species of interest that meets these criteria would be $^{86}$Sr.  $^{86}$Sr is a boson (for fermionic species, the contact interaction vanishes for atoms of the same spin state in different motional states of the same well, so our protocol would not be applicable without somehow using multiple spin states), and has an unusually high s-wave scattering length of roughly 800 times the Bohr radius \cite{de2008two}, leading to large interaction shifts $U_{01}$.   Strontium also has a narrow-linewidth optical transition with a linewidth of 7.5~kHz, suitable for ground-state sideband cooling \cite{norcia2018microscopic, cooper2018alkaline}.  Other bosonic species could be used as well, provided that strong enough interactions can be achieved, for example with magnetic Feshbach resonances, and that ground-state cooling is possible, for example using Raman sideband cooling.

In practice, the optical potentials could be implemented in several ways, with appropriate modifications depending on the variant of the protocol.  In any case, we assume the atoms to be confined in a three-dimensional optical lattice potential, whose depth can be tuned independently in all three directions.  This defines the ``base" potential.  The ability to tune the lattice depths independently enables one to increase the confinement along directions where tunneling is not required, which increases the interaction strength for two particles occupying the same site.  For the different protocols, we assume that the appropriate number of planes within the base potential are initially loaded, or that atoms remaining in undesired planes can later be removed.  On top of the base potential is superimposed an auxiliary potential, whose form depends on the variant of the protocol to be implemented.  

For loading target sites within a two-dimensional plane, the auxiliary potential could simply consist of tightly focused optical tweezers, incident from a direction orthogonal to the plane and projected through a high-numerical aperture optical system.  These serve both to create the required offset between target sites and neighboring reservoir sites, and by choosing the size of the tweezer spot, can also push down the potential barriers to enable faster tunneling.  

For loading target planes, the role of the auxiliary potential is simply to shift the energy of the target plane relative to its neighbors, and potentially of those neighbors relative to their outward neighbors if initial shift operations are applied to increase the reservoir density.
Practically, this could be implemented by superimposing a long-wavelength standing-wave auxiliary lattice formed from two beam intersecting a small angle onto the base lattice.  The planes of the base lattice that experience the lowest energy potential from the auxiliary lattice define the target planes.  Favorable initial conditions could be achieved by loading a single plane of a variable-wavelength ``accordion" auxiliary lattice at long wavelength, then transferring these atoms into several planes of the base lattice prior to ground-state cooling and parity projection on individual sites.  The ratio of the wavelengths of the auxiliary and base lattices then defines how many planes of atoms the target plane has available to draw atoms from.  

Finally, for the shuffle protocol, the auxiliary potential consists of a non-sinusoidal potential with twice the period of the base potential along the tunneling direction, and variable relative phase to the base potential.  The auxiliary potential could be created either by projecting a pattern onto the lattice using a high numerical aperture optical system, or by combining two polarizations of light in a bowtie-configuration lattice \cite{sebby2006lattice}.  In either case, the wells of the auxiliary potential can be aligned to the base potential with a phase such that they create both an offset between selected neighboring wells, and to push down the potential barrier between the wells to facilitate tunneling \cite{sebby2006lattice}.  In addition to these two periodic potentials, we assume an additional potential to be present that creates a wall on the left side of the array (for example by pushing a specific potential well off resonance), and a linear potential gradient applied along lattice to avoid additional unwanted resonances.

\section{Estimates of experimental parameters}\label{sec:app_params}
Here, we provide estimates for the relevant parameters achievable in a realistic experimental system.  As a benchmark case, we consider $^{86}$Sr atoms confined within a base lattice formed by retro-reflection of 813~nm light, as this is the ``magic" wavelength that causes zero differential shift to the strontium clock transition.  While this feature is not required for our protocol, it makes it a desirable wavelength for other applications of a strontium system.

\noindent \textbf{Tunneling rate:}
The tunneling rate $J_{01}$  between the ground state of one well and the first excited state of its neighbor is a critical quantity for the success of this protocol. Analytical expressions exist for the tunneling rates within a given band of a uniform lattice, but to our knowledge not for our nonuniform inter-band situation.  We thus calculate the tunnelling rate numerically for our desired potential landscape by simply integrateing the Schr\"{o}dinger equation.  We find the ratio of $J_{01}$ to $J_{00}$ to be approximately four for depths of the base lattice of at least 10~$E_R$, where $E_R = \hbar^2 k^2/2 m$.  Here, $k$ is the wavenumber of the light used to form the lattice, and $m$ is the mass of the atoms.  An analytical expression exists for $J_{00}$ \cite{Bloch2008}:

\begin{equation}
    J_{00} \simeq \frac{4}{\sqrt{\pi}}E_R(\frac{V}{E_R})^{3/4}\mathrm{exp}[-2(\frac{V}{E_R})^{1/2}] 
\end{equation}
Where $V$ is the depth of the lattice.  
For our benchmark system with a lattice depth of $V = 13 E_R$ (corresponding to a harmonic oscillator frequency of $\omega = 2 \pi \times 23$~kHz), we calculate a tunneling rate $J_{01}\simeq 2 \pi \times 175$~Hz.  This can easily be made slower by increasing the depth of the lattice.  In principle, it could be made faster as well in a shallower lattice, but after applying the required shift to neighboring wells, the potential becomes substantially distorted at the energy of the occupied states, and our expressions are no longer valid.  

\noindent \textbf{Interaction shifts:}
The interaction shifts that prevent double occupation of sites in our protocols result from the contact interaction between atoms in the first motional excited state and the ground state.  These shifts (under the assumption of unperturbed eigenstates) are approximated by:
\begin{equation}
    U_{01} = \frac{1}{2}\sqrt{\frac{8}{\pi}}k a E_R (\frac{V_x}{E_R})^{1/4}(\frac{V_y}{E_R})^{1/4}(\frac{V_z}{E_R})^{1/4}
\end{equation}
where $V_i$ represents the depth of the lattice in the $i$th direction.  The factor of $1/2$ multiplying this expression comes from the finite overlap between the ground state and first excited state wavefunctions ($U_{00}$ is a factor of 2 larger).  

For our benchmark system (a = 800 $a_0$) in cases where the tunneling occurs along a single direction (shuffle and layer protocols), depths of $V_x, \ V_y, \ V_z$ = 13,~30,~30~$E_R$ give an interaction shift $U_{01} = 2 \pi \times 10$~kHz.  When simultaneous tunneling in two dimensions is desired, the same interaction shift can be achieved using $V_x, \ V_y, \ V_z$ = 13,~13,~70~$E_R$.  For systems with more typical scattering lengths of $a \simeq 100$~a$_0$, the lower interaction shifts could be at least partially compensated by increasing the depth of the lattices in directions orthogonal to tunneling.

\noindent \textbf{Tunneling fidelities:}
In principle, our protocols could be implemented purely with resonant tunneling -- the depths of neighboring wells could be quickly brought into the desired resonance condition, then held there for a time $\pi/J_{01}$.  This approach would have the advantage of speed, but suffers from a high sensitivity to the relative depths of the neighboring wells, and requires precisely timed operations.  For total depths of the optical potential near 100~$E_R$ (accounting for the tight confinement required in non-tunneling directions in order to ensure large interaction shifts), this would require neighboring wells to be balanced at the part-per-thousand level in total lattice depth.  Because most of the intensity is associated with the base-potential, which can be created using retro-reflected lasers, this may not be unreasonable.  Roughly 10\% of the total potential in this scenario would be contributed by the auxiliary potential, so this would have to be controlled at the 1\% level.  The protocols will be substantially more robust however if an adiabatic ramp through resonance is used, so we assume this condition in all subsequent analysis.  

We numerically calculate the probability of an atom tunneling either into an empty well, or into an occupied well (represented by a shift of $U_{01} = 2 \pi \times 10$~kHz, as estimated above).  For a linear sweep over a range of $10 J_{01}$, centered about resonance, and over a duration $2\pi \times 5/J_{01}$ (roughly 30~ms for a tunneling rate $J_{01}\simeq 2 \pi \times 175$~Hz), we expect a transfer fraction of approximately 95\% into the empty well, and below $2\times 10^{-4}$ for the occupied well.  Such ramps only require control of the total potential at the 1\% level, and of the auxiliary potential at the 10\% level.  


To compare with prior work, tunneling fidelities at the 99\% level or higher have been demonstrated in the ground band of systems with optical superlattices \cite{lohse2016thouless, yang2020cooling, koepsell2020robust}.  Given the additional requirements of speed and tunneling between bands for this proposal, we think that fidelities above 90\% would be readily achievable in practice, and would not present a major performance limitation.  

\begin{figure}[!htb]
\includegraphics[width=\linewidth, ]{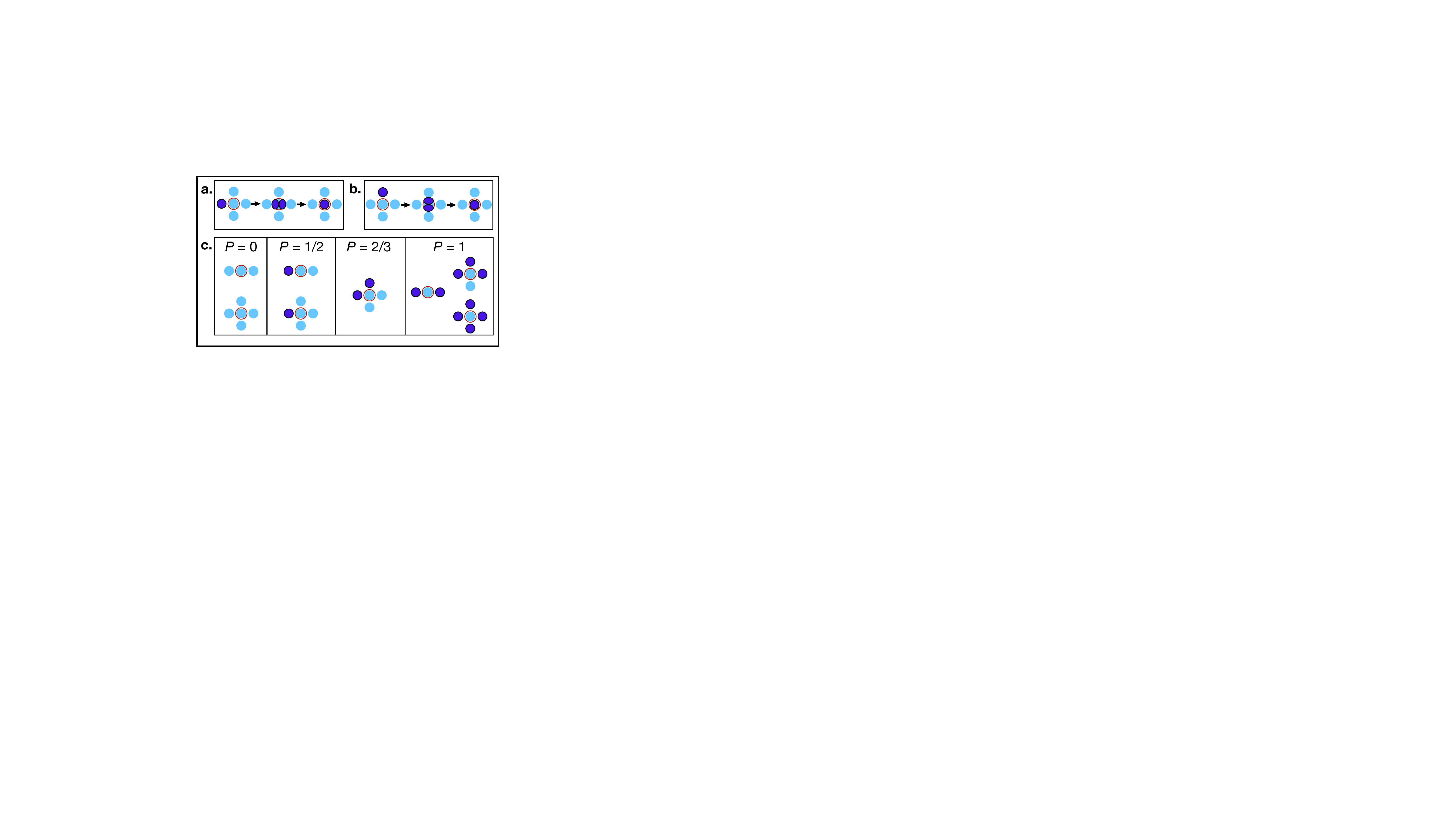}
\caption{Tunneling properties within nearest-neighbor subsystem.  \textbf{a., b.} The motional excited state into which the atoms tunnel is two-fold degenerate, with a node either along the vertical (a.) or horizontal (b.) axis.  Atoms tunneling into the target site populate one of these two states, depending on whether they tunneled horizontally or vertically.  During the tunneling process, an atom is thus confined to its original row or column, but can tunnel through the target site to the reservoir site directly opposite (not shown) \textbf{c.} The probability of a target site being filled depends on the number and configuration of filled neighboring reservoir sites.  Here we show the unique filling possibilities (omitting those that are equivalent up to reflection or rotation of the system), along with the associated probability for filling the target site, assuming a perfect adiabatic ramp of the target site depth.  }
\label{fig:filling}
\end{figure}

\noindent \textbf{Atom loss:}
Atom loss is the core limitation for our protocol.  Atom loss can occur either directly, though collisions with background gas, or as a result of our rearrangement protocol, through light-assisted collisions with other atoms.  The latter occurs if two atoms occupy the same site, which could be caused either by imperfect tunneling suppression from of the interaction shift, or through errors in ground state cooling.  

The effect of background gas collisions simply depends on the duration of the shift cycle relative to the vacuum lifetime in the experiment.  For our predicted tunneling rates of several hundred Hz, and associated ramp times of a few 10's of ms, we expect each cycle of the rearrangement to take well below 100~ms, as ground state cooling and changes in trap configuration for the cooling should be possible on a timescale of several milliseconds.  
As atomic lifetimes of 100 seconds or more are now somewhat common, we expect this loss to be at the part per thousand level per cycle, and could in principle be better if resonant tunneling is used instead of adiabatic ramps.  

The impact of imperfect ground-state cooling can have several impacts on our protocol.  If an atom in a target site begins a shift cycle in the n=1 state, it may tunnel into a neighboring empty reservoir state, leading to an effect similar to loss of the atom.  
More concerning would be the possibility of an atom tunneling into a well occupied by a neighbor, enabled by an accidental resonance between higher motional states (whose spacing becomes rapidly nonlinear with increasing N due to the relatively shallow potentials required for appreciable tunneling rates) and causing a double-occupancy.  The tendency for this to happen would likely depend sensitively on the exact optical potential used, but we consider the somewhat pessimistic case that imperfect ground state cooling leads to the loss of the non-cooled atom.  The fundamental limit on the performance of ground state cooling is given by $\bar{n} = 5(\gamma/\omega)^2/16$, where $\bar{n}$ represents the average number of motional quanta along a given direction, $\gamma$ is the decay linewidth used for cooling, and $\omega$ is the trap frequency \cite{wineland1987laser}.  This expression assumes cooling from three directions in an isotropic trap.  For realistic trap frequencies of $\omega = 2 \pi \times 100$~kHz during cooling, this implies that motional excitation should be limited to the level of a few parts per thousand.  

As described above, we find from simulation that the chances of an atom tunneling onto an occupied lattice site in a single rearrangement cycle can be below $2 \times 10^4$.  Because this would lead to the correlated loss of both atoms, the effect of this loss differs by protocol.  In the case of the shuffling protocol, both lost atoms are from the target state, so the effect on the final filling fraction is larger than single-atom loss.  However, because the loss is correlated, if we are only interested in the probability of preparing a perfect array, the effect is similar to loss of a single atom.  In the target sites and target planes protocols, only one of the atoms is from the target site, so the loss of the second atom is far less consequential.  

In principle, it may be possible to avoid the loss associated with double-occupation of lattice sites, as loss occurs only as a result of subsequent light-assisted collisions.  However, this would likely lead to a permanent double-occupation of the site in the final configuration, so for simplicity, we assume here that light-assisted collisions occur after each shuffling step (either through deliberate application of photoassociation light or as a byproduct of ground state cooling) and causes the loss of both atoms.

While the loss rates from double occupation may be acceptable for these parameters, they could be improved in several ways.  First, because of the quadratic scaling of the double occupation probability with tunneling rate, a relatively minor reduction in tunneling rate (deeper lattice) could dramatically reduce these errors, though at the expense of longer cycle times and greater sensitivity to lattice inhomogeneity.  Further, it may be possible to modify the light-assisted collision step to preferentially eject only a single atom, for example by using light blue-detuned light \cite{Grunzweig2010, lester2015, brown2019gray}.  

Adding these effects together, and acknowledging that they are rough estimates that will vary depending on the details of experimental implementation, these predictions indicate that a loss per cycle of less than 1\% should be possible, so we use this value as a benchmark for estimating the likely performance of our protocols.

\section{Simulating rearrangement protocols:}\label{sec:app_sims}
A fully quantum simulation of our protocols on a system of interesting size would be very challenging, due to the exponentially large Hilbert space.  However, because a dissipative step is present between each cycle of tunneling, we do not expect long-range entanglement to play a significant role.  Thus, to estimate the performance of our protocols, we perform simulations of the quantum dynamics for small sub-systems, from which we extract parameters that are used as inputs for classical simulations of larger systems.  

In the case of the shuffle protocol, the necessary parameters for the classical simulation are simply the tunneling fidelity $F$, and the loss rate per cycle.  The means for estimating these are described above.  

For the target sites protocol, the concept of the tunneling fidelity becomes a bit more complicated.  Each target site has four nearest neighbors. (For simplicity, we neglect diagonal neighbors, as the tunneling rates to these sites would be much lower, and these sites may be have different potential offsets from the nearest neighbors, depending on how the optical potential is generated.)  Each of the four nearest neighbors can either be occupied or empty, which leads to six distinct configurations of atoms (assuming symmetry of the neighboring sites): for all atom numbers except for two, there is a single unique configuration.  Two neighboring atoms can either be across the target site from each other or at a diagonal.  For each of these six configurations, we calculate the probability of the target site becoming occupied (tabulated in Fig.~\ref{fig:filling}), by integrating the Schr\"{o}dinger equation for the appropriate adiabatic ramp.  These values are then used in the classical simulation, as the base probability for filling a given target site, with the assumed imperfect fidelity multiplying this probability.  

We can gain intuition into the results using simple arguments.  The excited states to which the atoms couple are anti-symmetric about their centers, and we assume that the two excitation directions are degenerate.  If an atom tunnels into the target site, the state it excites only allows it to tunnel back into its original site or into the site across the target site from its original location.  Thus, atoms are confined to a given row or column.  At the beginning of the adiabatic ramp, we assume that the first motional excited states of the target site are higher in energy than the reservoir sites and that interactions are repulsive, so that the initial state always consists of a superposition of degenerate ground states, where the atoms are delocalized within their respective rows or columns.  After the ramp, the states involving one atom in the target site are lower energy than those that do not.  By comparing the degeneracy of the lowest energy states at the beginning  of the ramp to the degeneracy of lowest energy states to which these couple to end of the ramp, we can find the probability of filling the target site.  We confirm the results through explicit simulation for the relevant configurations.  


With the relevant parameters available, our simulations proceed by initializing randomly filled arrays of appropriate geometry, and then proceeding in a step-wise manner according to the relevant protocol.  Steps representing the coherent shift operations are interleaved with steps that mix the reservoir atoms (in the case of the target sites and layer filling protocols), and with steps representing atom loss.  For the shuffle protocol, each cycle consists of either a shift applied to even or odd pairs.  For the target sites protocol, we implement the reservoir mixing step in the following manner: the population of each reservoir site (iterated through sequentially) is swapped with one of its neighbors, chosen at random.  For the target layer protocol, we simply randomize the position of the atoms in each layer during each mixing step.  These implementations are chosen for simplicity --- we find that altering the details of the mixing step does not have a significant effect on the performance of the protocols, provided it is sufficiently randomizing.  This is especially true for predicting the maximum filling fraction.  

When filling target sites within a two-dimensional plane, we assume for simplicity that each target site interacts independently with its neighbors -- the probability of filling each site (randomly selected) is calculated in turn, with the atomic configuration updated in between.  This ensures that reservoir atoms are not double-counted, while maintaining computational simplicity.  If this situation is desired in practice, it could easily be achieved by offsetting the depths of the auxiliary potential applied to neighboring subsets of the target sites.

\section{Removing excess atoms:}\label{sec:app_removal}
All three of our protocols require the removal of excess atoms at the end of the filling cycles.  There would be several ways of accomplishing this, depending on the protocol.  

For the shuffling protocol, one option would be to use the same light that defined the first edge of the filled region to define the second, after shuffling is complete.   If the wavelength or polarization of this light is chosen such that it creates a differential shift on the cooling transition, then the cooling effect in these sites could be turned to heating.  Subsequent cooling light would then cause their removal from the trap.  Other options would include directing resonant light onto the region containing atoms to be removed, leading to heating or enabling future photo-ionization, or by applying light with strong spatial gradients, whose modulation at multiples of the trapping frequencies could lead to heating. 

For the target sites protocol, one could simply leave on the auxiliary potential that defines the target states while ramping off the lattice in the orthogonal direction (though retaining confinement perpendicular to the array), leaving the background atoms free to expand away.  Slight heating or spatial potential gradients could assist this process.  

For the target layer protocol, one could take advantage of the fact that the target plane lies at a minimum of the auxiliary potential, and thus at a point of zero (or low) gradient.  In this case, applying an intensity modulation at an odd harmonic of the trap frequency perpendicular to the target plane should selectively heat atoms in planes other than the target plane.  These heated atoms could be subsequently removed by lowering the lattice depth.  Alternatively, as in the case of the shuffle protocol, the auxiliary potential could be tuned (by a rotation of polarization for example) to provide a differential shift to either the cooling transition or an optical clock transition, enabling selective heating or shelving of atoms by layer.  Finally, charge-pumping techniques using optical superlattices\cite{koepsell2020robust} could be adapted to move atoms away from the target plane, facilitating their later removal or making them irrelevant to future operations.  

\end{document}